\documentclass[prd,amsmath,superscriptaddress,twocolumn]{revtex4}
\usepackage{graphicx}
\usepackage{dcolumn}   
\usepackage{hyperref}
\usepackage{color}
\usepackage{verbatim}
\newcommand{\smoothnu}{$\mathrm{smooth}\nu\mathrm{CDM}$}
\newcommand{\deltasmoothnu}{\delta_{\mathrm{smooth}\nu\mathrm{CDM}}}
\newcommand{\smoothnumodel}{\mathrm{smooth}\nu\mathrm{CDM}}
\newcommand{\Psmoothnu}{P_{cb}}
\newcommand{\lcdm}{$\Lambda$CDM}
\newcommand{\nulcdm}{$\nu\Lambda$CDM}
\newcommand{\nuede}{$\nu$DDE}
\newcommand{\deltacb}{\delta_{cb}}

\newcommand{\be}{\begin{equation}}
\newcommand{\ee}{\end{equation}}
\newcommand{\beqn}{\begin{eqnarray}}
\newcommand{\eeqn}{\end{eqnarray}}
\newcommand{\mnras}{MNRAS}

\begin{document}
\title{Effects of Massive Neutrinos and Dynamical Dark Energy on the Cluster Mass Function}
\author{Rahul Biswas}
\affiliation{HEP Division, Argonne National Laboratory, 9700 S. Cass Avenue, Lemont, IL 60439, U.S.A}
\affiliation{eScience Institute and the Department of Astronomy, University of Washington, 3910 15th Ave NE, Seattle, WA 91591, U.S.A}
\affiliation{The Oskar Klein Centre for CosmoParticle Physics, Department of Physics, Stockholm University, AlbaNova, Stockholm SE-10691}
\author{Katrin Heitmann}
\affiliation{HEP Division, Argonne National Laboratory, 9700 S. Cass Avenue, Lemont, IL 60439, U.S.A}
\affiliation{MCS Division, Argonne National Laboratory, 9700 S. Cass Avenue, Lemont, IL 60439, U.S.A}
\author{Salman Habib}
\affiliation{HEP Division, Argonne National Laboratory, 9700 S. Cass Avenue, Lemont, IL 60439, U.S.A}
\affiliation{MCS Division, Argonne National Laboratory, 9700 S. Cass Avenue, Lemont, IL 60439, U.S.A}
\author{Amol Upadhye}
\affiliation{HEP Division, Argonne National Laboratory, 9700 S. Cass Avenue, Lemont, IL 60439, U.S.A}
\affiliation{School of Physics, The University of New South Wales, Sydney NSW 
2052, Australia}
\author{Adrian Pope}
\affiliation{CPS Division, Argonne National Laboratory, Lemont, IL 60439, USA}
\author{Nicholas Frontiere}
\affiliation{HEP Division, Argonne National Laboratory, 9700 S. Cass Avenue, Lemont, IL 60439, U.S.A}
\affiliation{Department of Physics, University of Chicago, Chicago, IL 60637, USA} 
\date{\today}
\begin{abstract}
The presence of massive neutrinos affects the growth of large-scale structure in the universe, leaving a potentially observable imprint on the abundance and properties of massive dark matter-dominated halos. Cosmological surveys detect large numbers of these halos in the form of rich groups and clusters, using the information as an input to constraining the properties of dark energy. We use a suite of N-body simulations that include the effects of massive neutrinos as well as of dynamical dark energy to study the properties of the mass function. As in our previous work, we follow an approach valid at low neutrino mass, where the neutrino overdensities are assumed to be too small to act as a significant nonlinear source term for gravity. We study how well a universal form for the halo mass function describes our numerical results, finding that the use of an appropriate linear power spectrum within the formalism
yields a good match to the simulation results, correctly accounting for the (neutrino mass-dependent) suppression of the mass function. 
\end{abstract}

\maketitle
\section{Introduction}
%
Currently available cosmological observations are well explained by the standard paradigm of the
\lcdm~model~\cite{2015arXiv150201589P, 2012MNRAS.427.3435A, 2014A&A...568A..22B} which presently consists of a cosmological constant, cold dark matter, a small fraction of baryonic matter, and an even smaller fraction of massless species including photons and neutrinos. Results from neutrino flavor oscillation experiments show that neutrinos, while light, are not massless: the difference of squared masses of neutrino species is at least $m_{21}^2 = (7.53 \pm 0.18)\cdot 10^{-5} \rm{eV}^2$~\cite{Tanabashi:2018oca}. This is the first extension of the neutrino sector in the Standard Model and it is possible that this sector is more complex (see, e.g., Ref.~\cite{Aguilar-Arevalo:2018gpe}). The most stringent non-cosmological upper limits on neutrino masses come from Tritium beta-decay experiments and constrain the electron neutrino mass $m_{\nu_e} \lesssim 2 \rm{eV}$ at the $95 \%$ limit~\cite{2008RPPh...71h6201O}. On the other hand, the most recent Planck Collaboration results~\cite{Aghanim:2018eyx} provide a $95\%$ upper bound on the sum of neutrino masses to be $\sum m_\nu \lesssim 0.24 \rm{eV}$ using the cosmic microwave background (CMB) temperature and polarization power spectrum along with measurements of CMB lensing. While these constraints are somewhat sensitive to the high-$l$ modeling of the polarization spectrum
and lensing of the CMB, the constraint on the sum of masses improves to $\sum m_\nu \lesssim 0.12 \rm{eV}$ when the Planck temperature and polarization anisotropy spectra are used in combination with baryon acoustic oscillation (BAO) data from BOSS~\cite{2012MNRAS.427.3435A}. Thus, the most competitive constraints on the masses of neutrinos at current times, and for the foreseeable future, are provided by cosmology.

Massive neutrinos alter the cosmic evolution and the large-scale structure distribution relative to a \lcdm~model, leaving an imprint on
 several observables at different length scales (for recent reviews, see~Refs.~\cite{2011ARNPS..61...69W, 2012arXiv1212.6154L, 2014arXiv1401.6085F}).  
Ongoing and future cosmological surveys such as the Dark Energy Survey (DES)~\cite{2005astro.ph.10346T}, the Dark Energy Spectroscopic Instrument (DESI)~\cite{2013arXiv1308.0847L}, the Large Synoptic Survey Telescope (LSST)~\cite{2009arXiv0912.0201L}, and
the Square Kilometer Array (SKA)~\cite{Bacon:2018dui}, which are designed to probe the accelerating universe, will also be sensitive to the presence of massive neutrinos. Such observations will potentially provide the tightest constraints on neutrino masses and may even be able to discriminate between the normal and inverted mass hierarchies~\cite{2014arXiv1401.6085F}. It has also been noted that some of the imprints on cosmological observables due to neutrino masses are degenerate with effects that can be attributed to different models of dark energy, as noted, for example, in Ref.~\cite{Wang:2005vr}.
Therefore, in order to constrain neutrino masses or dark energy from survey data, it is important to study the behavior of the relevant observables, simultaneously allowing variations in dark energy parameters and neutrino properties. 

Future observations will combine measurements of the CMB anisotropies with probes of fluctuations of the matter density at different scales and redshifts, typically measured using two-point statistics (e.g., BAO, weak gravitational lensing).
In addition, the measured abundance of late-forming and relatively rare objects like galaxy groups or clusters provides very useful information related to dark energy. This information is conveniently encoded in the number density of halos as a function of mass at a particular redshift, i.e., the halo mass function. Due to the dynamics of massive neutrinos, replacing a fraction of dark matter by massive neutrinos in the cosmological matter budget, leads to a suppression of the mass function, reducing the numbers of clusters for a given variance in density of fluctuations. We will study this effect in detail in this paper, using cosmological simulations.

In the context of a \lcdm~model or a $w$CDM model, where the equation of state of dark energy is characterized by a model parameter, $w$, constant in time, the requirements for calculating precise and accurate cluster mass functions have been studied extensively. Historically, the cluster mass function was first estimated by Press and Schechter~\cite{1974ApJ...187..425P} in an Einstein de-Sitter model through an ansatz that mapped the linear matter power spectrum to the mass function through a redshift independent mapping, with parameters obtained from spherical collapse. A key characteristic of the Press-Schechter approach is universality: aside from background quantities like densities, only the mass variance as a function of smoothing scale at the redshift of interest ($\sigma(R, z)$ at different $R$) is required to estimate the mass function. In order to provide a good match from the predictions to results from N-body simulations of multiple cosmological models, the mappings required modifications which led to significantly more accurate fits~\cite{2001MNRAS.323....1S, 2001MNRAS.321..372J} that still respect universality. Further detailed studies extending to $w$CDM models use redshift dependent mappings~\cite{2010MNRAS.403.1353C, 2011ApJ...732..122B} to match the results of N-body simulations, mildly breaking universality at low redshifts. The study of multiple $w$CDM models in Ref.~\cite{2011ApJ...732..122B} demonstrated that calculating the spherical collapse critical density parameter, $\delta_c$, for the cosmological model in question does not improve the match of the mass function fits to the simulation results (see also Ref.~\cite{2001MNRAS.321..372J}). It was shown that semi-analytic fits match N-body results across cosmologies only at the 10-15\% level of accuracy, leading to efforts to develop emulators~\cite{2010ApJ...713.1322L, 2016ApJ...820..108H} to predict the halo mass function for different cosmologies.

While progress on providing predictions for the mass function has been made for cosmologies including massive neutrinos both in terms of semi-analytic work~\cite{Ichiki:2011ue,2014PhRvD..89f3502L, 2014PhRvD..90h3518L,2014PhRvD..90h3530L} as well as estimates based on N-body simulations~\cite{2010JCAP...09..014B, Castorina:2013wga, 2013MNRAS.428.3375A,Castorina:2015bma},
detailed studies 
in the context of $w$CDM models including neutrinos have not been carried out.  Obtaining good estimates of the cluster mass function is challenging, requiring high statistics
 of halo abundance, with large numbers of halo particles, along with other requirements (see, e.g., Ref.~\cite{2011ApJ...732..122B}). Adding neutrinos, a form of ``hot'' dark matter, can cause a number of numerical problems if not done carefully. Because of their small masses, neutrinos have high thermal velocities at early times, leading to problems in particle simulations involving neutrinos as a separate species. Additionally, 
 the low mass of the neutrino tracer particles compared to the CDM tracer particles can create unphysical scattering-induced segregation effects. 
 
These problems are exacerbated for small neutrino masses, like the ones indicated by the recent cosmological data from Planck. Recent efforts to counter this usually involve approximations of some form, such as starting the simulations at a later stage~\cite{2010JCAP...09..014B}, or adding a linear power spectrum from neutrinos~\cite{2013MNRAS.428.3375A}.
The method for including the effects of massive neutrinos in dynamical dark energy N-body simulations that we use here follows our earlier work in Ref.~\cite{2014PhRvD..89j3515U}; as further discussed in Section~\ref{sec:method} below, this method allows us to bypass many of the issues that make including neutrinos problematic in other schemes.   

The paper is organized as follows. In Section~\ref{sec:method} we first give a brief introduction describing the effects of neutrinos on the large-scale structure distribution. Next, we discuss our implementation
of massive neutrinos in the simulation code HACC (Hardware/Hybrid Accelerated Cosmology Code)
and introduce our approximate treatment of neutrinos. In Section~\ref{sec:mfu} we provide a discussion of universality
and how the concept is employed in our mass function fitting approach. We then describe the set-up of our simulations in Section~\ref{sec:sims}. In Section~\ref{sec:results} we show results from our N-body simulations and investigate the validity of universality for three different cosmologies. We provide conclusions and an outlook in Section~\ref{conclusions}.   

\section{Background and Methodology}
\label{sec:method}
In this Section we describe how the dynamical effects of massive neutrinos are taken into account in our approach. We first discuss the basic physics and then present the numerical implementation.

\subsection{Dynamical Effects of Massive Neutrinos in Structure Formation}
In the early universe, neutrinos are in thermal equilibrium and follow an 
ultra-relativistic Fermi-Dirac distribution. As the universe expands, the interaction rate of neutrinos drops, resulting in neutrinos falling out of thermal equilibrium. The evolution of their momenta are dictated by the conditions at this point and scale as the inverse of the scale factor $\sim 1/a$. Therefore, even at late times the velocity distribution may be obtained from an approximate ultra-relativistic Fermi-Dirac distribution
 with an effective temperature $T \sim 1/a$, despite the typical neutrino velocities being small  $v \sim 150 \mathrm{km/s} \left(1{\rm eV}/m_\nu\right) (1.0 +z).$ 
 
The thermal velocities erase the growth of neutrino perturbations at scales below the free streaming scale $k_{fs} \propto {a H}/{v}$, but allow its growth at large scales. This is in contrast to cold dark matter, where the thermal dispersion is close to zero unless sourced by structures.
Since density fluctuations source the growth of matter perturbations, this implies a stronger 
growth of matter fluctuations at large scales than small scales. Thus, in 
a universe where a fraction of the matter density includes significant contributions from massive 
neutrinos, the growth of CDM fluctuations becomes scale-dependent.

While neutrinos source the growth of structure only on large scales where their density is relatively smooth, their velocities mostly prevent them from being captured by the gravitational potential wells of dark matter structures. We employ a simple argument, following the discussion in Refs.~\cite{Ichiki:2011ue,2013MNRAS.428.3375A}, to obtain an estimate of the comoving size of dark matter structures that can bind neutrinos of a certain mass. Since a neutrino is captured by a potential well if the well grows during the neutrino's travel time through it -- and the time-scale of growth of cosmic structures is related to a Hubble time $1/H(t)$ -- only those neutrinos that are slow enough to not cross the structure in a time larger than the time required for the potential to grow can possibly be captured. This leads to an estimate that neutrinos of mass $m_\nu \sim$ 1eV might be weakly captured by structures of comoving size $\sim 1$Mpc, i.e., a cluster-sized object. Given the constraints from Planck, a single neutrino is likely to be much lighter and even more weakly bound to a cluster. A different analysis based on studying spherical collapse models \cite{2014PhRvD..89f3502L} also suggests that for neutrino masses much lighter than one eV, neutrinos would not be significantly captured even by a cluster. At the same time we also know that high mass objects such as groups or clusters are relatively rare, with most of the significant matter overdensities being present in the form of much smaller halos. 

Based on the above discussion, our treatment of massive neutrinos includes the following effects (discussed here in the specific context of \nulcdm~where the dark energy component is assumed to be arising from a cosmological constant):
\begin{enumerate}
\item {Background evolution of the universe:} Due to the addition of
a neutrino background, the overall mass density changes. For the larger neutrino masses
that we will be concerned with, the background density of neutrinos calculated
from the Fermi-Dirac distribution for late times of interest $z \leq 200,$ 
evolves like matter:
\be
\rho_\nu \sim (1 + z )^{-3}.
\ee
\item {Initial conditions:} The presence of the neutrino fluctuations 
implies that the spectrum of CDM and baryon density fluctuations at an earlier 
time (such as $z =100$) is different from that of a $\Lambda$CDM model with 
the same background cosmology and variance of fluctuations, as given by $\sigma_8$. 

\item {Scale-dependent clustering:} Due to the free streaming of neutrinos, the growth of the matter power spectrum is scale dependent, as the cold dark matter is also sourced by neutrino fluctuations at large scales, but not at small scales.
We can describe this at the linear level via
\be
\delta \rho^{cb}_{\nu\Lambda CDM}(k,z) = D(k,z) \delta \rho^{cb}_{\nu\Lambda CDM}(k,z = 0).
\label{eqn:physicalnucdm}
\ee
\end{enumerate}
Note that neutrino capture in dark matter dominated halos is most likely insignificant as discussed above and is not included in our numerical approach.

\subsection{Simulation Methodology}
The evolution of cold collisionless matter fluctuations in an expanding universe is well-described by the Vlasov-Poisson equation (VPE). Because of its high dimensionality and the formation of structure at very small scales that can cause breakdown of grid-based techniques, N-body methods are the standard approach to solve the VPE. The finite value of the particle mass used in such simulations can lead to artificial collisional effects that must be controlled on the spatial scales of interest. In the case of a multi-species simulation, collisionality induced segregation effects can have potentially serious effects, especially if the relative tracer particle mass ratios are large (for a recent study, see Ref.~\cite{emberson}). This is of particular concern for massive neutrinos, since the mass ratio between the masses of particles representing CDM-baryons $m_{cb}^p$ and neutrinos $m^p_\nu$, can be very large, roughly two orders of magnitude. (Representing a neutrino (CDM-baryon) mixed dark matter model with 
particle implementations of neutrinos requires the relation
$m^p_{\nu}/m^p_{cb} = (N_{cb}/N_{\nu})(\rho_{\nu}/\rho_{cb})$
between the masses assigned to the simulation particles and their numbers.)
Another problem is the large velocity dispersion of the neutrinos relative to the coherent initial velocity perturbation, motivating methods to suppress the associated shot noise, some of which require many more neutrino tracer particles. However, this is not a necessary condition (see, e.g., Ref.~\cite{Banerjee:2016zaa}). To suppress various N-body artifacts in \nulcdm~N-body simulations (for an early paper, see Ref.~\cite{1999ApJ...524..510G}), it is also customary to turn on the gravitational interaction for neutrinos only at relatively low redshifts ($z<10$, e.g., Ref.~\cite{Inman2015}).

To avoid these difficulties, and to enable large volume simulations with good mass resolution needed for group and cluster-scale mass function studies, we use a different approximation~\cite{2014PhRvD..89j3515U}. This approach works particularly well at low neutrino masses, as it is a leading order expansion in $f_{\nu}=\Omega_{\nu}/\Omega_m$ in the spirit of Ref.~\cite{Saito:2009ah} (see also Ref.~\cite{Agarwal:2010mt}). The associated neutrino mass range is nicely consistent with the upper limits from Planck. 

For any cosmological model, the sum of neutrino masses $\sum m_\nu$ is related to important background quantities through
\be
\rho_\nu (z)  = \rho_{crit}(z=0)  \frac{\sum m_\nu}{94 \mathrm{eV} h^2}(1 +z)^3,
\label{eqn:rhonuevol}
\ee
to evolve the neutrino density and solve the Friedmann equations for the background values of all constituents. Eqn.~(\ref{eqn:rhonuevol}) is obtained from the Fermi-Dirac distribution and valid at low redshifts, $z \lesssim 200$, which are relevant for the simulation.
The particles in the N-body simulations only represent cold dark matter and baryonic particles (gravity-only, no gas dynamics or feedback), and not neutrinos. Neutrinos are taken to be smooth and have no density fluctuation at any scales, and are evolved just like, for example, the energy density corresponding to a cosmological constant (except with a null equation of state parameter). Thus the particles have a mass
$$m_p = \rho_{crit}(z=0) \left(\Omega_c + \Omega_b\right) 
\frac{\mathrm{SimVol}}{N_p}$$
and this avoids the problems associated with treating neutrinos as a separate species in the N-body simulations.
However, since the fluctuations on large scales are not being sourced by the neutrinos, starting with a particle distribution having the linear power spectrum calculated for the cosmological model at the starting redshift will result in a $z=0$ power spectrum that has less power on large scales than the linear matter power spectrum for the same cosmological model. Therefore, 
we initialize the simulation to have a compensated power spectrum at the starting time that has slightly higher power at larger scales than the linear power spectrum. More specifically, we use the following prescription: 
\begin{enumerate}
\item First, we calculate the $z=0$ power spectrum associated with baryons and cold dark matter only for the cosmological model with massive neutrinos:
$P_{\nu\rm{LCDM},lin}^{cb}(k, z=0)$. This is done by using the $z=0$ transfer functions for the cold dark matter and baryonic components calculated from CAMB~\cite{Lewis:1999bs,Lewis:2002nc} and the primordial power spectrum.  
\item Next, we solve the ODE for the linearized growth of perturbations to obtain the growth function in a universe where the neutrino density has zero fluctuations to obtain $D^2_{smooth\nu}(z)$ and find the compensated linear power spectrum in the $\rm{smooth}\nu$ model as 
\be 
P_{\mathrm{smooth}\nu,lin}^{cb} (z) = D^2_{\mathrm{smooth}\nu}(z) P_{\nu \mathrm{LCDM},lin}^{cb}(z=0)
\ee
\item Finally, the initial conditions of the particles in the N-body simulation are set by initial displacements and 
velocities of each particle from a uniform grid, where these displacements and velocities are calculated from a Zel'dovich move so that they are a realization
drawn from the power spectrum $P_{\mathrm{smooth}\nu,lin}^{cb} (z)$.
\end{enumerate}
This means that for any redshift, the linear cold dark matter, baryon component of the power spectrum is related to the true linear power spectrum with the same components as:
$$
P_{\mathrm{smooth}\nu}^{cb} (z) = \frac{D^2_{\mathrm{smooth}\nu}(z)}
{D^2_{\nu\mathrm{LCDM}}(k,z)}P_{\nu \mathrm{LCDM}, lin}^{cb}(z),
$$
where $D^2_{\nu\mathrm{LCDM}}(k,z)$ is a scale dependent growth function in the $\nu$LCDM model. Thus,
$P_{\mathrm{smooth}\nu}^{cb} (z)$ is always larger than 
$P_{\nu \mathrm{LCDM}}^{cb}(z)$ at large scales. Physically, this means that the initial conditions have more clustering at large scales, but since the power on these scales does not grow as fast as it would if the neutrinos acted as a source, the power spectrum catches up to $P_{\nu \mathrm{LCDM}, lin}^{cb}(z)$ at $z=0$. 

In Ref.~\cite{2014PhRvD..89j3515U}, we have shown that the 
power spectrum obtained from this set of simulations correctly matches 
linear perturbation theory at the largest scales at low redshifts ($z < 2$),
as well as the power spectrum calculated using Time-RG perturbation
theory at mildly nonlinear scales. We can use time-RG perturbation theory to check whether our methodology is consistent, by 1) checking the agreement at scales before the perturbation theory breaks down, and 2) checking that the change in the CDM-baryon power spectrum is small when the 
sourcing due to the clustered neutrinos is included and when it is not. As shown in Ref.~\cite{2014PhRvD..89j3515U}, both of these tests are satisfied to high accuracy over the redshift range of observational interest. Additionally, a test carried out recently comparing the Mira-Titan Universe emulator~\cite{lawrence17} $P(k)$ prediction with $\sum m_{\nu}=0.15 \rm{eV}$, which uses the same method as described here, against a low-noise N-body simulation with massive neutrinos (albeit at worse mass resolution than the simulations described here) shows very good agreement in the nonlinear power spectrum over the expected range of scales~\cite{Banerjee:2016zaa,Banerjee:2018bxy}.

All of these results provide an excellent confirmation of our basic methodology, which, as stated above is particularly well suited to large-scale structure simulations in large volumes, that evolve large particle numbers, and include a lower range of neutrino masses. Finally, note that we are working in the ``single species approximation" rather than with, for example, three degenerate species. This assumption is expected to be a good approximation at low neutrino mass.

\section{Mass Function Universality}
\label{sec:mfu}
The main idea of universality is that halo mass functions, while arising from the 
nonlinear regime of growth of large-scale structure, can nevertheless be estimated entirely from knowing the linear power spectrum for any cosmological model. The mass function is written as
\be
\frac{dn}{d\ln{(M)}}  = \frac{\rho_{bg}}{M} f\left[\sigma (M,z)\right]
\frac{d\ln{[\sigma^{-1}(M,z)]}}{d\ln{(M)}},
\label{eqn:universalfit}
\ee
where the values of $\sigma(M,z)$ are calculated from the mass of the halos 
and the comoving density of components clustering in the halos:
\be
M = \frac{4\pi}{3} R^3(M) \rho_{bg} (z = 0), \qquad \sigma(M,z)  = \sigma[R(M), z],
\ee
where $\sigma(R(M), z)$ is given by
\be
\sigma^2(R, z)  = \int dk k^2 \tilde{W}^2(k) P(k,z) \left(2\pi^2h^3/h^3\right),
\label{eqn:sigmaR}
\ee
and $W$ is a tophat filter in real space and $P(k,z)$ is the linear matter power spectrum. 

The idea behind universality was implicit in the Press-Schechter like studies 
 which suggested forms for $f(\sigma)$ based on quantities from 
spherical collapse. More recently, N-body simulations have been
used to demonstrate the (approximate) validity of universality by reproducing the mass function
of a single cosmological model at different redshifts by using the values
 of $\sigma(R)$ at those redshifts~\cite{2001MNRAS.321..372J}. This confirms the idea in the form
that the linear power spectrum at different redshifts provides sufficient
information to estimate the mass function, but it does not 
show that the same universal form can estimate the mass function to a sufficiently high accuracy
 for different models (and for different definitions of halo mass). Studies focusing on $w\mathrm{CDM}$ models show that the assumption of universality
holds only at the $10-15\%$ level. Thus, it is unclear just how well this description will work for a {\nulcdm} model or a dynamical dark energy model.   

Usually the linear power spectrum and the variance 
of density fluctuations are quantitative measures of the clustering in a 
single time snapshot of the universe, and have no dynamical connection. In contrast, inserting these quantities into the universal fitting 
functions through $\sigma(R, z),$ which can provide estimates of the time 
evolution of observables like the mass function, uses them as dynamical 
quantities. Thus the exact convention and definition used for the power
spectrum is extremely important. For an Einstein de-Sitter model, the 
choice of power spectrum is obvious, and is the power spectrum of 
of the overdensity $\delta(x) \equiv {\delta \rho(x)}/{\rho}.$
For a {\lcdm}~model, the power spectrum used is the power spectrum of 
the matter overdensity 
\be
\delta_m(x) = \frac{\delta \rho_m(x)}{\rho_m(x)}. 
\label{eqn:deltam}
\ee
For a model with neutrinos, several possibilities exist to generalize the choice of the power spectrum for universality studies from Einstein de-Sitter: first, one may include all background components that evolve as  $\sim (1+z)^3$ in calculating $\delta_m(x)$ as defined above. Evolving perturbations linearly then yields the familiar density contrast that is used in the computation of power spectra in linear perturbation theory (as, e.g., in CAMB)~ defined as
\be
\delta_{lin}(x) = \frac{\delta\rho_c(x,z) + \delta\rho_b(x,z) +\delta\rho_{\nu}(x, z)}{\rho_c (z)+ \rho_b (z) + \rho_\nu (z)}.
\label{eqn:delta_lin}
\ee

A second choice in generalization is to use the 
overdensity of the clustering CDM-baryon components and leaving the neutrino density out in analogy to how dark energy density is treated, recognizing that neutrinos hardly cluster at the relevant scales. This gives us: 
\be
\deltacb(x) = \frac{\delta\rho_c(x,z) + \delta\rho_b(x,z)}{\rho_c (z)+ \rho_b (z)}.
\label{eqn:deltacb}
\ee
Finally, we investigate the matter overdensity 
for the {\smoothnu} model, using the idea of Eqn.~(\ref{eqn:delta_lin}) to calculate the background density, but explicitly setting the spatial variation of the neutrino density to zero:
\be
\deltasmoothnu(x) = \frac{\delta\rho_c(x,z) + \delta\rho_b(x,z)}{\rho_c (z)+ \rho_b (z) + \rho_\nu (z)},
\label{eqn:deltasmoothnu}
\ee
which is consistent with the fact that neutrinos behave like CDM on very large scales. We note that Eqn.~(\ref{eqn:deltacb}) and Eqn.~(\ref{eqn:deltasmoothnu}) differ only by a time dependent normalization but this is important for universality studies due to the nonlinearities of the fitting functions [Eqns.~(\ref{eqn:universalfit})].

\begin{table*}
\begin{center}
\caption{Cosmological models investigated in this paper; each simulation evolves
$3200^3$ particles in a (2.1~Gpc)$^3$ volume.
\label{tab:cosmoparams}}
\begin{tabular}{ccccccccccc}
Model &  $m_p [10^{10}$M$_\odot]$ & $\omega_{cdm}$ & $\omega_b$ & $\omega_{\nu}$ & $n_s$ &  $\sigma_8$ & $h$ & $w_0$ & $w_a$ & $\Sigma m_\nu$ [eV] \\ \hline
$\Lambda$CDM & 1.05$$ &0.1109 & 0.02258 & 0.0 & 0.963 & 0.800 & 0.7100 &-1.0 & 0.0 & 0.0 \\
$\nu\Lambda$CDM & 0.97 & 0.1009 &  0.02258 & 0.010 & 0.963 & 0.800 & 0.7100 &-1.0 & 0.0 & 0.94\\
\nuede  & 1.19 & 0.1280 & 0.0232 & 0.00311 & 0.880 & 0.805 & 0.7342 &-1.2 & -1.11 & 0.292 \\
\end{tabular}
\end{center}
\end{table*}

\section{Simulation Specifications}
\label{sec:sims}

A sufficiently high precision study of the mass function of halos requires 
large volume N-Body simulations with good mass and spatio-temporal resolution. The requirements
for such studies, and the systematic biases incurred if the requirements
 are not met, have been examined in detail in a number of previous papers, see, e.g. Refs.~\cite{Warren:2005ey,Lukic:2007fc,2010MNRAS.403.1353C,2011ApJ...732..122B,McClintock:2018uyf}.
Due to the steepness of the halo mass function, a small fractional bias in the halo mass
manifests itself as a much larger shift in the mass function. In the following we summarize the main requirements for gravity-only simulations to provide accurate results for mass function measurements.
\begin{enumerate}
\item Large simulation volume: This is necessary to obtain good statistics for high 
mass halos, which are rare because of the steep fall-off of the mass function at high masses. Additionally, boxes that are too small suppress nonlinear evolution and consequently depress the mass function.  
\item High mass resolution: Each halo needs to be sampled with a sufficient number of particles to obtain stable estimates of the halo mass at the smallest mass scale of interest. 
\item Force resolution and time-stepping: The force resolution needs to be matched to the mass resolution and should be sufficiently small compared to the spatial extent of the halo. Time-stepping errors can also lead to a diffusive effect and need to be properly controlled. 
\end{enumerate}

Our work is based on a set of simulations  using the HACC framework. The details of the implementation of HACC for 
$\Lambda$CDM simulations are described in Ref.~\cite{Habib:2014uxa, 2012arXiv1211.4864H,2009JPhCS.180a2019H}, while the 
modifications implemented to take into account effects from neutrinos and dynamical dark energy
equation of state are given in Ref.~\cite{2014PhRvD..89j3515U}.
As detailed in the following, we have made careful choices for our simulation set up to ensure that we accurately address the listed conditions. Each simulation evolves $3200^3$ particles in a (2100~Mpc)$^3$ box.
This volume leads to good statistics in the cluster mass range up to halo masses of $\sim 10^{15}$M$_\odot$. Depending on the exact cosmological parameters, the particle mass in the simulations is approximately  $m_p \sim 10^{10}$M$_\odot$, which provides good particle sampling in the halo mass range characteristic of groups and clusters. Our studies cover a mass range of $\sim 4.2\cdot 10^{12}$M$_\odot$ to $\sim 1.4\cdot 10^{15}$M$_\odot$, leading to a particle sampling of $\sim 420-140,000$ per halo. As was first shown in Ref.~\cite{Warren:2005ey} and later confirmed but slightly revised in Ref.~\cite{2011ApJ...732..122B}, the sampling for the low mass halos in our simulations will lead to a mass uncertainty of less than 2\% and for the high mass halos less than 0.1\%. The force resolution in the simulations was set to 6.6kpc. Following the tests carried out in Ref.~\cite{2011ApJ...732..122B}, this force resolution ensures accurate capture of the halos at the mass ranges considered here. Time stepping tests were carried out in the past for HACC and we ensured that our simulations are properly converged with the settings chosen. Some of these simulations were used in Ref.~\cite{2014PhRvD..89j3515U} to study the validity of the Time-RG approach to predict the matter power spectrum. The tests there showed excellent agreement of the simulations with the Time-RG predictions, confirming the accuracy of our simulations on large scales. The initial redshift is chosen to be $z_{\rm in} = 200$, ensuring that the halos had enough time to evolve correctly. Simulation outputs at $z=0,1,2$ are used to study the evolution of the mass function, bracketing a range of interest relevant to observations. 

We report results from simulations covering 1) a reference $\Lambda$CDM model, 2) a \nulcdm~model with a single-species massive neutrino (with $\sum m_{\nu}=0.94 \rm{eV}$), and 3) a dynamical dark energy model, \nuede,~with a lower neutrino mass sum. Cosmological parameters and the resulting mass resolution for each model are given in Table~\ref{tab:cosmoparams}; note that spatial flatness is assumed. More details about the models are discussed in Section~\ref{sec:results}.

\section{Results}
\label{sec:results}

Our studies of the impact of neutrino masses on the halo mass function use the following cosmological models (the parameters of the models are 
summarized in Table~\ref{tab:cosmoparams}): a reference flat \lcdm~model with 
massless neutrinos for comparison and test purposes, a \lcdm~ 
model with a single-species massive neutrino where 
the energy density of massive neutrinos accounts for a nontrivial fraction 
($7.5\%$) of the matter density, or in other words with $\sum m_{\nu}=0.94 \rm{eV}$ ($\nu\Lambda$CDM 
in Table~\ref{tab:cosmoparams}), and a dynamical dark energy simulation with a lower contribution from neutrino mass, $\sum m_{\nu}=0.292 \rm{eV}$ (or where the energy density of the neutrinos is $3.2\%$ of the matter density, {\nuede} in Table~\ref{tab:cosmoparams}). The dynamical dark energy  with equation of state is parameterized via $w_0, w_a$, following Refs.~\cite{chevalier,linder03}: $w(a)=w_0+w_a(1-a)$. The density of the cold dark matter in the $\nu\Lambda$CDM~model is lowered by
the amount of the massive neutrino density, so that the baryon and dark energy
densities is the same as in the \lcdm~model with massless neutrinos.
Both models are normalized to the same $\sigma_8$ (the variance of density
fluctuations smoothed on scales of $8h^{-1}$Mpc), leading to a larger amplitude 
for the power spectra for the $\nu\Lambda$CDM~model at large scales.
The neutrino masses are chosen to be large enough to cause distinguishable variations in the mass function from a corresponding model with massless neutrinos and a similar value of $\sigma_8,$ and be relevant for the range of masses directly probed by observation. The \nuede~model is an example of less studied models which involve both massive neutrinos and phantom dark energy models with $w_{0}=-1.2$~ and $w_{a}=-1.11$.  

\subsection{Mass Function Measurements}
\label{sec:massmeas}

First, we briefly describe the methodology and mass definition used to measure the halo mass function from our simulations. The halos are obtained from outputs at three redshift slices 
$z = 0, 1, 2$ using the friends-of-friends (FOF) algorithm~\cite{davis} with a linking length of $b = 0.2$. This value for the linking length is customary for mass function studies involving universality, see, e.g., Refs.~\citep{2001MNRAS.321..372J,Lukic:2007fc,More:2011dc}. We then bin the halos in the mass range 
from $3.0\cdot 10^{12}-1.0\cdot 10^{15} h^{-1}$M$_\odot$ into $20$ equal mass 
bins of size $\Delta M$, each represented by a corrected arithmetic mean 
$\bar{M}$ of the masses of halos in that bin. Next, we find the number of halos 
$N_{bin}$ in each bin. In evaluating the mean mass $\bar{M},$ 
the masses of individual halos are adjusted using the empirical correction factor 
discussed in Refs.~\cite{Warren:2005ey,2011ApJ...732..122B} which accounts for the FOF mass bias due to the finite number of particles in halos. 
The correction is given by $n^{\rm corr}_h=n_h(1-n^{−0.65}_h)$, where $n_h$ denotes the number of particles in the halo.
We only report mass function results in a 
mass range where the number of particles in halos is larger than 400 and the correction factor is therefore small. The mass function $dn/d\ln{(M)}$ is estimated by $(N_{bin}/\rm{SimVol})(\bar{M}/\Delta M),$ 
where $\rm{SimVol}$ is the comoving volume of the simulation. The statistical 
uncertainty in these calculated values of the mass function is due to 
Poisson fluctuation of the number of halos in a given mass bin.

\begin{figure}
\begin{center}
\includegraphics[width=8.9cm]{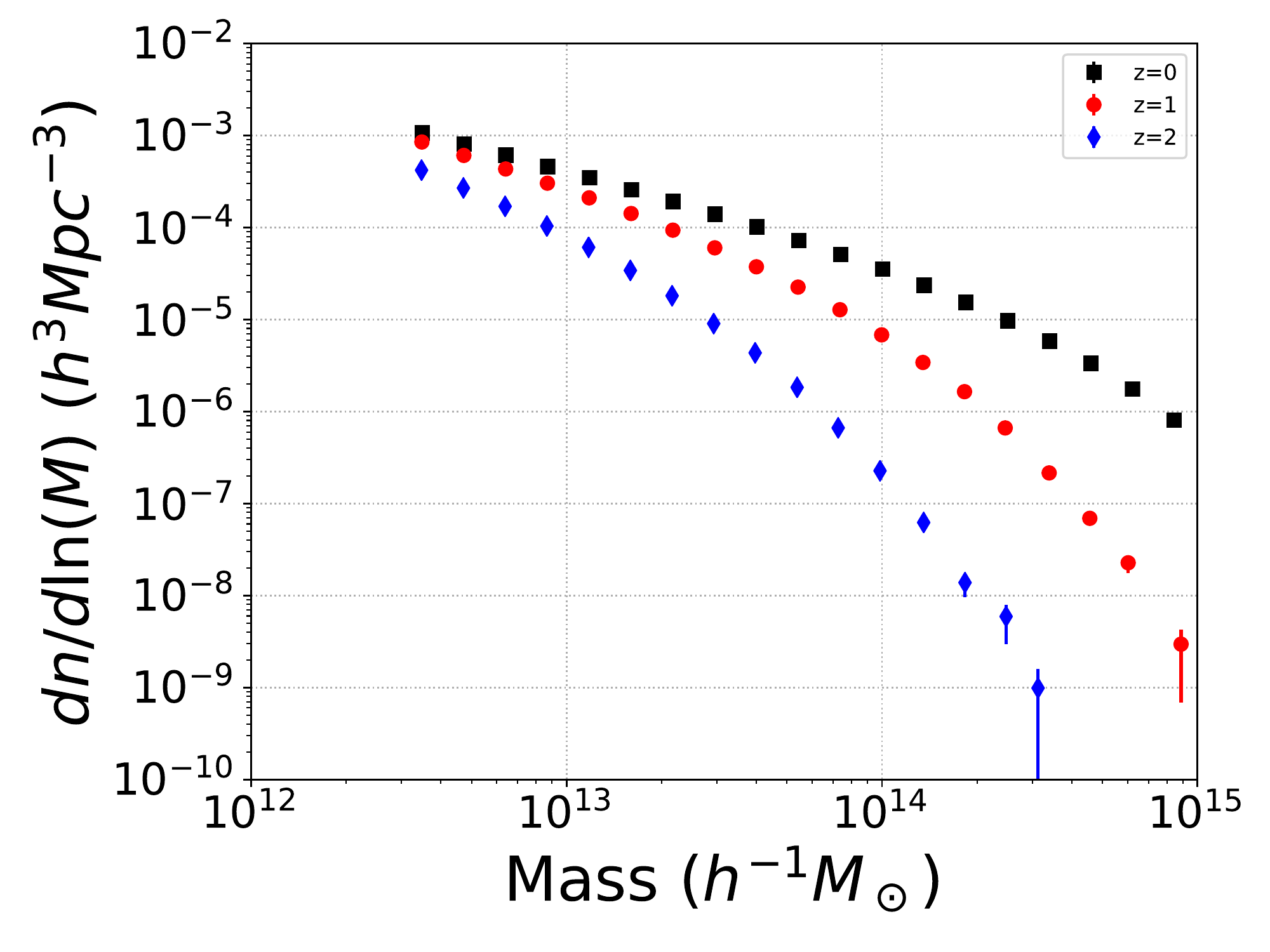}
\caption{Halo mass function at redshifts $z= 0, 1, 2$ with Poisson errors from the \lcdm~cosmology simulation 
(see Table~\ref{tab:cosmoparams} for details of the parameters).}
\label{fig:M000_mf}
\end{center}
\end{figure}

\begin{figure}
\begin{center}
\includegraphics[width=8.9cm]{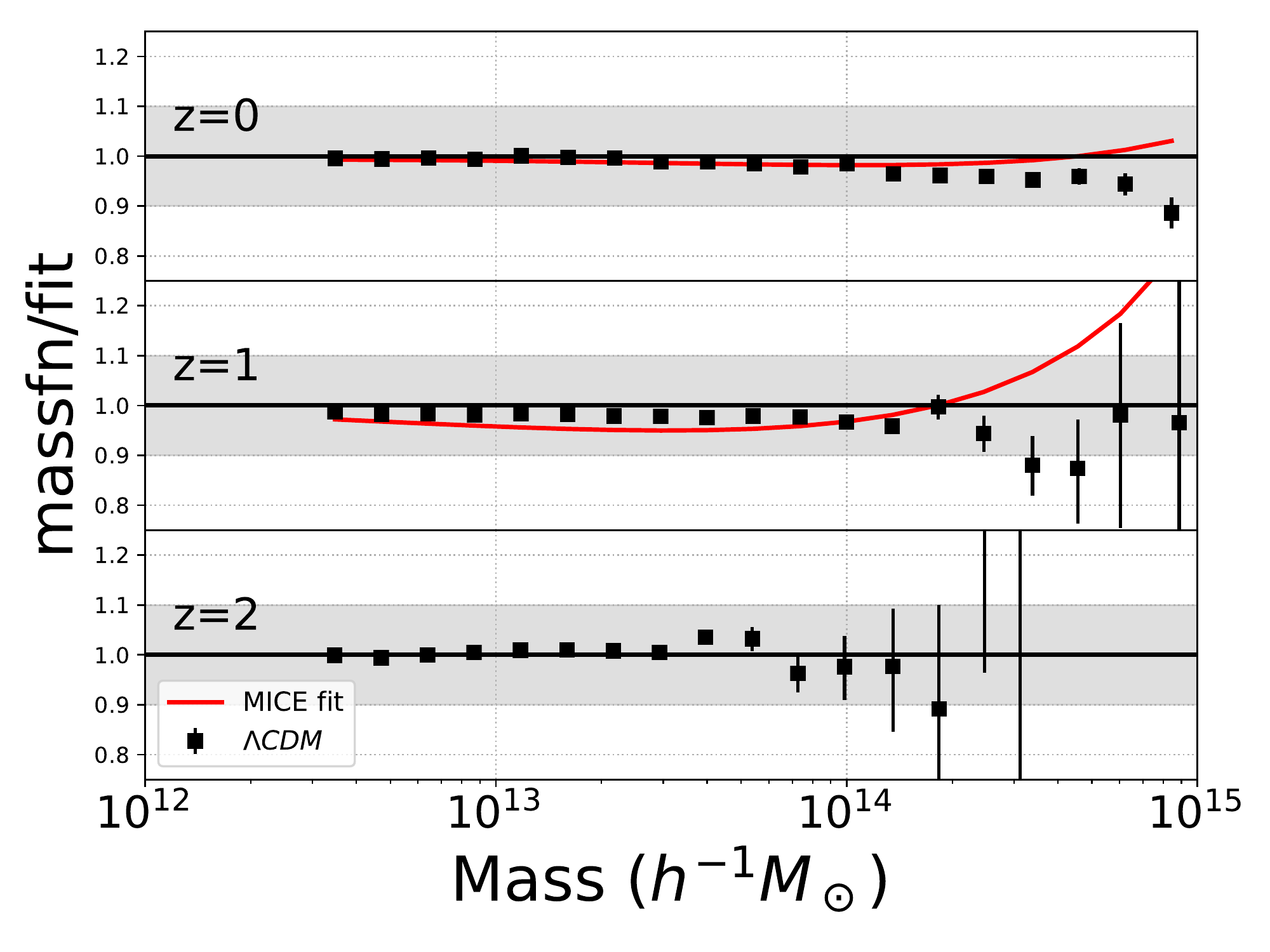}
\caption{Comparison of the $\Lambda$CDM mass function results (Fig.~\ref{fig:M000_mf}) to two fitting functions: Black squares with error bars show the 
ratio of the simulation results to the predictions from Bhattacharya et al.~\cite{2011ApJ...732..122B} taken as the reference. 
The red line represents the ratio of the MICE fit~\cite{2010MNRAS.403.1353C} to the reference. 
The redshift range for the MICE fit is restricted [0,1], hence the ratio is not shown for $z = 2.0$.
The gray band represents a $\pm 10\%$ variation from unity.}
\label{fig:M000_umf}
\end{center}
\end{figure}

\subsection{$\Lambda$CDM Mass Function}

We first investigate the halo mass function for the \lcdm~cosmology with zero neutrino mass. 
The mass function (FOF, $b=0.2$) for such models has been the subject of several studies, therefore we can easily test for consistency with previous work.
In Fig.~\ref{fig:M000_mf}, we show the halo mass function at three different redshifts. The asymmetric error bars $\sigma_{\pm} =  \sqrt{N_{bin} + \frac{1}{4}} \pm \frac{1}{2}$ represent uncertainties due to Poisson fluctuations and are computed from the 
number of halos in the bin $N_{bin}$, following Ref.~\cite{fnal_recommendation:possion}.

Our results for the $\Lambda$CDM model should be consistent with those obtained from previous N-body 
simulations at good accuracy. Since other studies used slightly different cosmological parameters, a direct comparison of the mass function is difficult. Instead, we compare our results to 
universal fitting functions in terms of $f(\sigma)$. These have been advertised to be valid over 
certain mass ranges and certain families of cosmology and therefore provide a good test. Throughout this paper, we use two different expressions for the fitting function. The first fitting function is provided in Ref.~\cite{2011ApJ...732..122B} and is of the form
\be
\label{eqn:bhattfit}
f(\sigma, z) = \tilde{A}\sqrt{\frac{2}{\pi}}
\exp{\left[-\frac{\tilde{a}\delta_c^2}{2\sigma^2}\right]}
\left(1 + (\frac{\sigma^2}{\tilde{a}\delta_c^2})^{\tilde{p}} \right)
\left(\frac{\delta_c \sqrt{\tilde{a}}}{\sigma}\right)^{\tilde{q}}.
\ee
Based on 
a large suite of simulations, this was estimated to be accurate at about two percent 
for $\Lambda$CDM models, in the mass range $10^{11}-10^{15} h^{-1}$M$_\odot$ 
at redshifts z = $0 - 2.$ For brevity, we will refer to this fit in the following as the Bhattacharya fit. The second fitting function we use for our investigation is provided in Ref.~\cite{2010MNRAS.403.1353C}:
\be
f(\sigma) = A (\sigma^{-a} + b ) \exp{( - c/\sigma^2 )}.
\label{eqn:micefit}
\ee
This fit is based on the MICE simulations, and was advertised to be correct at the ten percent
level over the mass range $10^{10}-10^{15} h^{-1}$M$_\odot$ with an associated redshift range of $z=0-1.$ We will refer to this fit as the MICE fit for the remainder of the paper.

We display our mass function results for the $\Lambda$CDM model at three redshifts in Fig.~\ref{fig:M000_mf}. In Fig.~\ref{fig:M000_umf} we compare our results with the two fits. We use the functional form for the mass function given in Eqn.~(\ref{eqn:universalfit}) and for the two fits insert the expressions for $f(\sigma)$ given in Eqs.~(\ref{eqn:bhattfit}) and (\ref{eqn:micefit}). We use the Bhattacharya fit as baseline in this plot.
The black squares show the ratio of the simulation result with respect to the Bhattacharya fit, while the red line shows the ratio of MICE fit and the Bhattacharya fit directly. The error bars represent the statistical errors from our simulations; the gray bands show the ten percent region of agreement with respect to the Bhattacharya fit. 

The $\Lambda$CDM results shown in Fig.~\ref{fig:M000_umf} are in good agreement
with the previous studies, keeping in mind that we have a single box, so some realization scatter is to be expected, in particular at high masses. We would need a much larger simulation volume to actually check if we are consistent with the Bhattacharya fit at the two percent level. The 
comparison with the MICE simulations shows good consistency at 
$z=0,$ but there is some divergence at redshift $z=1$. It should
be noted, however, that this is at the very edge of the redshift range over which
the MICE simulations were originally fitted. 

\begin{figure}
\begin{center}
\includegraphics[width = 9cm]{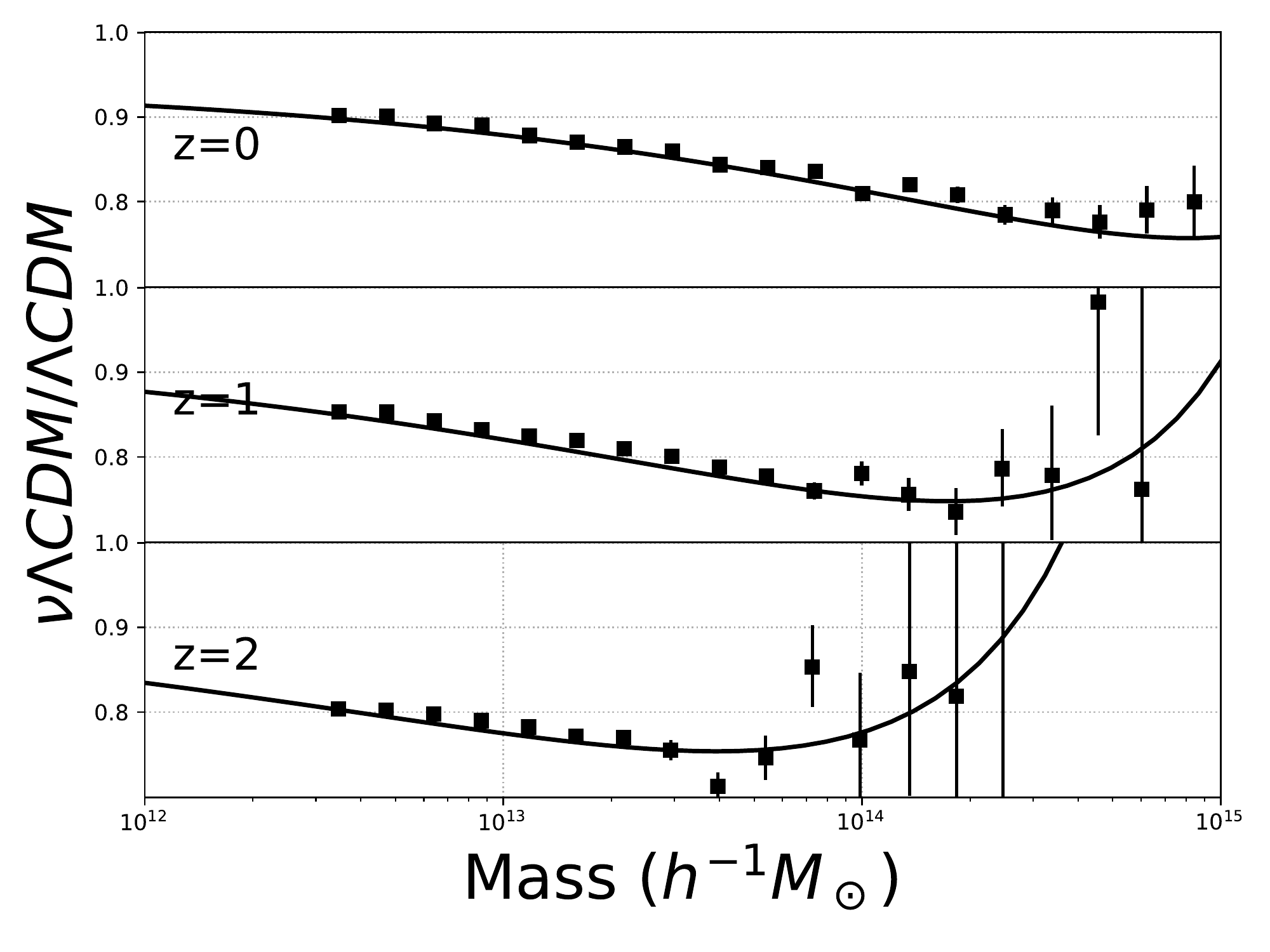}
\caption{Halo mass function ratio (black squares) at redshifts $z=0, 1, 2$ from the $\nu\Lambda$CDM and \lcdm~models (Table~\ref{tab:cosmoparams}) showing suppression of the mass function by massive neutrinos. In the $\nu\Lambda$CDM run, a fraction of cold dark matter in the \lcdm~model is replaced by massive neutrinos with $\sum m_\nu = 0.94 \rm{eV}$, while holding $\sigma_8$ unchanged. The solid black curve shows the predicted fitting function ratios taken from Ref.~\cite{2011ApJ...732..122B}, modified to use Eqn.~(\ref{eqn:deltasmoothnu}) in the universality fit as discussed in Section~\ref{sec:mfu}. As a cautionary note we point out that results at high masses at $z=1, 2$ are uncertain due to limited statistics, both for the simulation results and for the asymptotics of the fitting form used.}
\label{fig:suppression}
\end{center}
\end{figure}

\subsection{Neutrino Effects on the Mass Function}

The verification of our mass function results for the $\Lambda$CDM model via comparison to previously obtained fitting functions now allows us to turn to the effects of massive neutrinos on the halo mass function. We begin by considering massive neutrinos in a $\Lambda$CDM cosmology. In what follows below, the mass of the halo is taken to be 
the mass of the FOF halo of CDM-baryon particles, while the background density
is taken to be the density of the CDM-baryons and neutrinos.

Fig.~\ref{fig:suppression} shows the suppression of the mass function 
due to a fraction of the cold dark matter component being replaced by massive neutrinos. We measure
the ratio of the mass function of the $\nu\Lambda$CDM model
to the mass function of the \lcdm~model (see Table~\ref{tab:cosmoparams} for the details). The only difference between
these two models is that $7.5\%$ of the dark matter is replaced by a non-clustering smooth neutrino fluid where we use a single-species massive neutrino model. We note that the suppression in the mass function 
is quite large, particularly at high masses where at redshift $z = 0,$ this can be a $\sim 20 \%$ effect.
At higher redshifts, the effect is systematically larger. (The very good statistical accuracy of our results implies that in future we can study the suppression of the mass function quite comfortably at lower neutrino masses.) We note that at the higher redshifts halos at the upper end of the mass range are rarer, and our results become statistically limited. In particular the shape of the suppression cannot be predicted well at masses (approximately) greater than $3\cdot 10^{14}h^{-1}$M$_\odot$ at $z=0$, $10^{14}h^{-1}$M$_\odot$, at $z=1$, and $3\cdot 10^{13}h^{-1}$M$_\odot$, at $z=2$. However, larger simulations can easily be used to extend this range.

It is interesting to note that the form of the suppression -- up to the mass ranges indicated above -- is very well described by the Bhattacharya fit using the overdensity $\deltasmoothnu$ given in  Eqn.~(\ref{eqn:deltasmoothnu}) as the appropriate input solid curve in Fig.~\ref{fig:suppression}). The asymptotic forms of analytic mass function fits suffer from the same statistical problems as the raw data fits at high mass, and the upturn in the ratio at high masses should not be over-interpreted.

\begin{figure}
\begin{center}
\includegraphics[width = 8.6cm]{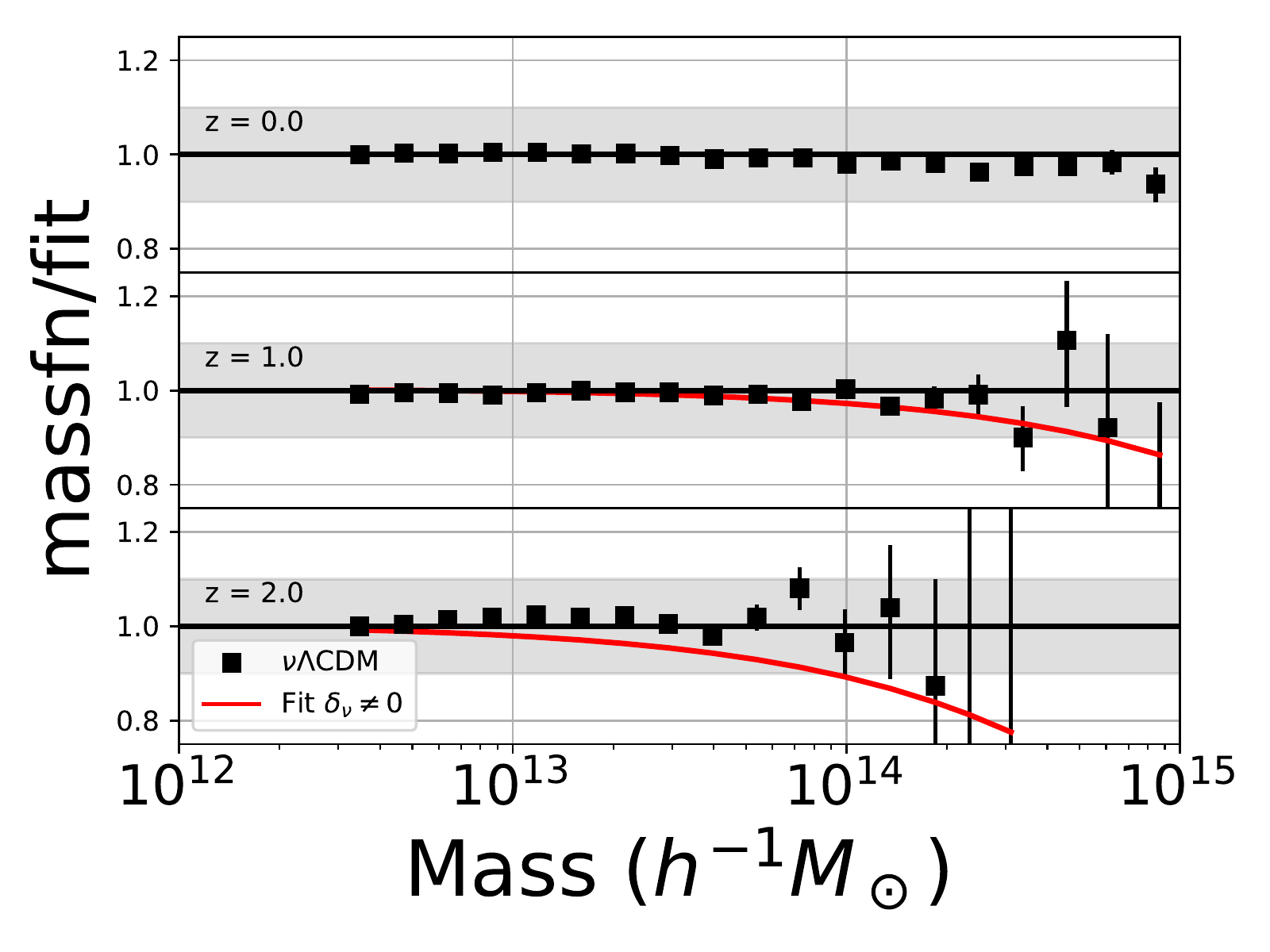}
\caption{Universality of the mass function in a $\nu\Lambda$CDM cosmology. The black
    squares show the ratio of the mass function evaluated from the simulations to the mass function predicted by the
universal fitting function form of Ref.~\cite{2011ApJ...732..122B} for the $\nu\Lambda$CDM model
cosmology ($\sum m_\nu = 0.94 \rm{eV}$) using $\Psmoothnu$ as the linear power 
spectrum used in calculating $\sigma(M)$ in the input to the fitting functions [Eqn.~(\ref{eqn:deltasmoothnu})].
The red curve shows the ratio of the fitting function with a linear power spectrum including the effects of neutrino sourcing leading to an effective 
scale-dependent growth [Eqn.~(\ref{eqn:delta_lin})]. The gray bands represent a 10\% error spread.
}
\label{fig:mfratio}
\end{center}
\end{figure}

To continue our investigation of how well universality-inspired fits work for the mass function in neutrino cosmologies, we present results for the \nulcdm~model alone in Fig.~\ref{fig:mfratio}. The primary purpose here is to show that the mass function from the simulation follows the universal form computed using the power spectrum  $P_{\smoothnumodel}$ based on the overdensity $\deltasmoothnu$ as given in  
Eqn.~(\ref{eqn:deltasmoothnu}), in contrast to the alternative option of using $\delta_{lin}$ as defined in Eqn.~(\ref{eqn:delta_lin}); the two predictions are the same at $z=0$, but will differ as a function of redshift. There is excellent agreement between the simulation results and the ``universal" prediction based on using $\deltasmoothnu$. In the statistically well-described region of halo masses, the fits with the linear power spectrum evaluated directly from CAMB, i.e., with the density contrast in Eqn.~(\ref{eqn:delta_lin}) used in the evaluations, appear to be mildly disfavored (red curves in Fig.~\ref{fig:mfratio}).

\begin{figure}
\begin{center}
\includegraphics[width = 8.9cm]{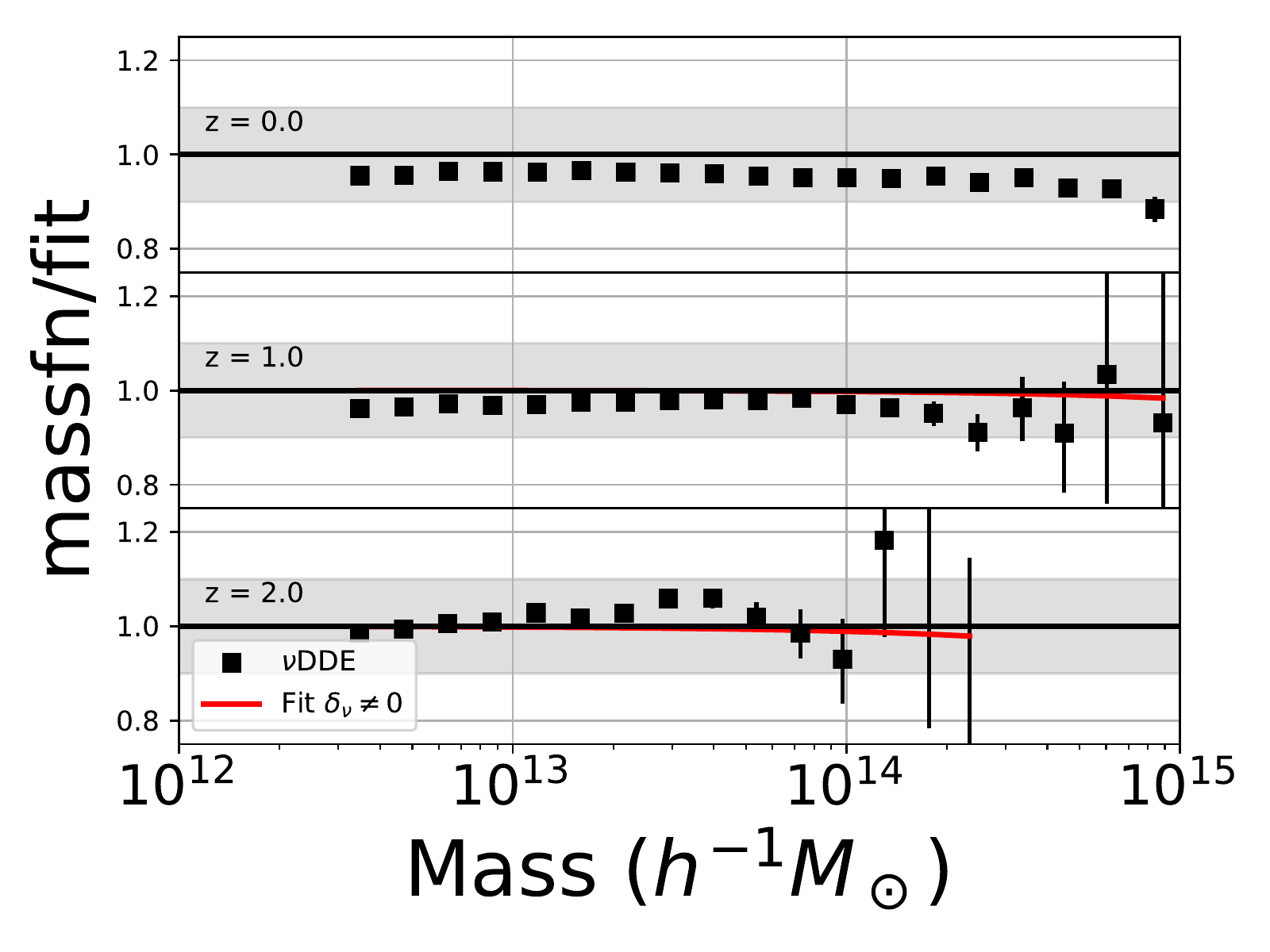}
\caption{Testing universality of the mass function in a cosmology with massive neutrinos and a dynamical dark energy (conventions as in Fig.~\ref{fig:mfratio}). The
universal fitting function of Ref.~\cite{2011ApJ...732..122B} is used for the 
$\nu$DDE model ($\sum m_\nu = 0.29 \rm{eV}$, in a dynamical dark energy cosmology) 
using $\Psmoothnu$ as the linear power 
spectrum used in calculating $\sigma(M)$ as input to the fitting functions.}
\label{fig:mfu_M011}
\end{center}
\end{figure}

\subsection{Neutrinos and Dynamical Dark Energy}

The final set of results is based on a cosmological simulation with massive neutrinos and a dynamical dark energy. This model has a smaller neutrino mass than the one used for the $\nu\Lambda$CDM case: $\Sigma m_\nu = 0.29$~eV along with equation of state parameters 
$w_0= -1.2$ and $w_a = -1.1$ (for the complete parameter set, see Table~\ref{tab:cosmoparams}). 

The results, shown in Fig.~\ref{fig:mfu_M011}, are presented in exactly the same way as in Fig.~\ref{fig:mfratio}, with the black squares showing the ratio between the mass function as measured from the simulation to the mass function obtained using the universal fitting functions with a smooth non-clustering neutrino component. In contrast, the red curves show the mass functions derived from universal fits using a linear power spectrum computed using the density contrast  of Eqn.~(\ref{eqn:delta_lin}) that evolves with redshift with effects of neutrino sourcing included. 
With a mass much smaller than $\sim 1$eV and in the range
that is closer to current observational constraints, the difference between these predictions is 
negligible. On the other hand, we note that with both massive neutrinos and dynamical dark energy with an equation of state much less than $-1$ driving it away from the normal \lcdm~models by causing aggressive acceleration,
universality does not hold as well in this case even at redshift $z=0$.

This result is consistent with the finding of Ref.~\cite{2011ApJ...732..122B} that accuracy of universal mass function predictions reduces to approximately $10\%$ for $w$CDM cosmologies. (Note, however, that this study did not include massive neutrinos.)
Neutrino masses within the 
current data bounds do not significantly increase the errors or significantly affect the 
universality violations of the mass function.  This is true for a 
cosmological constant as well as dark energy with a rapidly-varying 
equation of state.  However, even a $10\%$ error is unacceptably large for 
upcoming surveys.  Reducing this error will require either a theoretical 
advance in quantifying the universality violation or the emulation of the 
dark-energy-dependence of the fitting function based on a suite of 
simulations.

\section{Summary and Discussion}
\label{conclusions}
The abundance of rich galaxy groups and clusters is a key cosmological probe based on large-scale structure formation. The power of this probe relies in turn on robust theoretical predictions of the halo mass function, which currently require running large N-Body simulations. For a cosmology with massive
neutrinos, and where one also has to include the effects of nontrivial dark energy models, standard N-Body approaches can be very computationally expensive. Therefore, it is important to study whether simplified approaches made possible by certain assumptions can be reasonably effective, both for simplifying the simulations and in providing successful fitting forms for the simulation results. (Regarding the latter, an approach based on the assumption of universality is particularly attractive.)

The approach to the N-body simulations used here makes the simplifying assumption that, in the currently observationally allowed mass range, neutrinos cluster weakly in comparison to cold dark matter, allowing for an asymmetric treatment of neutrino perturbations. In previous work, the nonlinear power spectrum from these simulations
has been shown to be accurate in the perturbative regime by comparing to Time-RG perturbation theory~\cite{2014PhRvD..89j3515U}. A recent comparison with a full N-body approach also shows good agreement~\cite{Banerjee:2018bxy}. In this paper, we use the simplified simulation technique to predict the halo mass function for massive neutrino cosmologies.  

We present the mass functions obtained from our simulations and compare them
against fitting functions obtained using notions of universality. Having taken requisite 
care so that the simulations have the quality characteristics required for good
estimates of mass functions, we first show that our mass functions (at different redshifts) obtained
for a \lcdm~model are in good agreement with previous simulations performed 
with different N-Body codes and by different groups. We then compare the mass 
functions obtained from our simulations to those from universal fitting functions in the case of massive neutrinos and dynamical dark energy.

The prescription for extending universality to \nulcdm ~models is unclear,
with a variety of possible options, as discussed in Sec.~\ref{sec:mfu}. Given that the neutrino clustering scale is so large, and previous work based on spherical collapse models, we argue for a prescription that should best describe the results from N-body simulations. This is one where the 
quantities in the fits are calculated from the power spectrum of an 
overdensity $\deltasmoothnu$ as defined in Eqn.~(\ref{eqn:deltasmoothnu}).

Studying two 
different cosmological models, a $\nu\Lambda$CDM~cosmology which has $\sum m_\nu = 0.94 \rm{eV}$ 
and a dynamical dark energy model (see Table~\ref{tab:cosmoparams}),
we show (Figs.~\ref{fig:mfratio} and~\ref{fig:mfu_M011}) that 
our simulation results in both of these cases are in good agreement with the universal fits carried out as described above. 
We also show that the universal fitting functions do not match the mass functions
from the simulations for the \nuede~model for a dynamical dark energy model far away from 
\lcdm~at precision levels comparable to the other cases, even at $z=0$. This is consistent with previous studies, and points to the necessity for emulators in place of fitting functions for precision cosmology calculations.
\section{Acknowledgements}
Work at Argonne National Laboratory was supported under U.S. Department of Energy
contract DE-AC02-06CH11357. This research used resources of the Argonne Leadership Computing Facility, which is a DOE Office of Science User Facility supported under Contract DE-AC02-06CH11357.  

RB acknowledges partial support from the Washington Research Foundation Fund for Innovation in Data-Intensive
Discovery and the Moore/Sloan Data Science Environments Project at the University of Washington.

\end{document}